# A methodology for formulating dynamical equations in analytical mechanics based on the principle of energy conservation

To cite this article: Yinqiu Zhou and Xiuming Wang 2022 *J. Phys. Commun.* **6** 035006

View the article online for updates and enhancements.

# Journal of Physics Communications

**PAPER**

# A methodology for formulating dynamical equations in analytical mechanics based on the principle of energy conservation

**Yinqiu Zhou**[1,2,3]  **and Xiuming Wang**[1,2,3,*]

1   State Key Laboratory of Acoustics, Institute of Acoustics, Chinese Academy of Sciences, Beijing 100190, People's Republic of China
2   School of Physics Sciences, University of Chinese Academy of Sciences, Beijing 101408, People's Republic of China
3   Beijing Engineering Research Center for Drilling and Exploration, Institute of Acoustics, Chinese Academy of Sciences, Beijing 100190, People's Republic of China
*   Author to whom any correspondence should be addressed

**E-mail: wangxm@mail.ioa.ac.cn**

## Abstract

In this work, a methodology is proposed for formulating general dynamical equations in mechanics under the umbrella of the principle of energy conservation. It is shown that Lagrange's equation, Hamilton's canonical equations, and Hamilton-Jacobi's equation are all formulated based on the principle of energy conservation with a simple energy conservation equation, i.e., the rate of kinetic and potential energy with time is equal to the rate of work with time done by external forces; while D'Alembert's principle is a special case of the law of the conservation of energy, with either the virtual displacements ('frozen' time) or the virtual displacement ('frozen' generalized coordinates). It is argued that all of the formulations for characterizing the dynamical behaviors of a system can be derived from the principle of energy conservation, and the principle of energy conservation is an underlying guide for constructing mechanics in a broad sense. The proposed methodology provides an efficient way to tackle the dynamical problems in general mechanics, including dissipation continuum systems, especially for those with multi-physical field interactions and couplings. It is pointed out that, on the contrary to the classical analytical mechanics, especially to existing Hamiltonian mechanics, the physics essences of Hamilton's variational principle, Lagrange's equation, and the Newtonian second law of motion, including their derivatives such as momentum and angular momentum conservations, are the consequences of the law of conservation of energy. In addition, our proposed methodology is easier to understand with clear physical meanings and can be used for explaining the existing mechanical principles or theorems. Finally, as an application example, the methodology is applied in fluid mechanics to derive Cauchy's first law of motion.

## 1. Introduction

Classical mechanics mainly consists of Newton's second law of motion, Lagrange's equation, and Hamilton's principle, from which the dynamical equations of a system can be formulated. There is no doubt that the results from each of the three frameworks are equivalent, and all of the formulations eventually abide by the law of conservation of energy, for example, in classical mechanics [1], fluid mechanics [2], relativistic mechanics [3], quantum mechanics [4], elastodynamics/acoustics [5], electrodynamics [6], thermodynamics [7], and the studies on multi-physical fields coupling [8, 9]. More detailed discussions on the relationship between invariants and symmetric quantities shown in Noether's theorems are thoroughly treated in [10]. In nearly all of the published monographs or textbooks, generalized energy Hamiltonian $H$ is not conserved in a sense of mechanical energy when for example, the Lagrangian function $L$ is with time $t$ being directly explicit. However, it seems that it is not clear in this case, how the mechanical energy interacts with work done by external forces should be expressed when $\partial L/\partial t$ doesn't vanish.

On the other hand, in a general sense, the law of the conservation of energy, wherever civilization reaches, has been accepted as a fully established truth and has received the widest applications in natural sciences [11, 12] although there is still a debate in conservation characterization in some specific areas [13]. Specifically, it is said that the conservation of mechanical energy is among the deepest and most penetrating principles, and all of the fundamental forces obey the first principle of physics, i.e., the principle of energy conservation [14]. However, ever since Lagrangian mechanics was established, energy centered method has been always dominated the physics world through the principles of least action [1, 15, 16], rather than the law of conservation of energy although it has been used to validate the established dynamical equations, including Newton's second law of motion, Lagrange's Equation, Hamilton's variational principle, and so on .

In almost all of the monographs, textbooks, and published articles, it is said that various kinds of energy conservation formulations are 'derived' from Newton's law, Lagrange's equation, or Hamilton's variational principle. Even in quantum mechanics, Ehrenfest's theorem, related to kinetic and potential energy conservation, is obtained by using Schrödinger's equation [17]. According to analytical mechanics, it is safe to say that the conversion between energy and work, or energy conservation in a broad sense, is the consequence of Newton's second law of motion, Lagrange's equation, or Hamilton's variational principle. For example, a detailed explanation is that, in a conservative system, the mechanical energy conservation is due to the homogeneity of the time domain; momentum conservation, the homogeneity of the spatial domain; while angular momentum conservation, due to isotropy of the spatial domain [1, 18]. Although momentum, angular momentum, and energy conservation laws are derived assuming Newton's laws, these conservation laws are fundamental in nature that apply well beyond the domain of applicability of Newtonian mechanics [16]. Especially, the general principle of conservation of energy is a law governing all the natural phenomena that are known to date. There is no known exception to this law [19].

Since the principle of energy conservation has been so widely accepted, and it is natural to ruminate that, rather than using Hamilton's variational principle, why don't we start from the first principle of physics, the principle of energy conservation to construct the basic principles of analytical mechanics? Is that possible to take the principle of energy conservation as an axiom to reconstruct the whole mechanics?

As is said in our earlier review, the mechanical energy conservation equation is derived from Newton's second law of motion, Lagrange's equation, or Hamilton's principle. However, the logic that a positive proposition holds doesn't mean that its inverse proposition holds either. In the last decades, there have been several works in tackling this question [20]. Birkhoff seems to be the first one to use mechanical energy conservation naturally to inverse Lagrange's forces [21], while Pars pointed out that, an insidious fallacy that has appeared many times in the history of dynamics is the deduction of Lagrange's equations from Jacobi's integral or energy conservation [22]. That means Birkhoff's treatments on generalized forces are not correct [22]. However, Wang [23] pointed out, if virtual displacements in d'Alembert's principle were replaced by velocities, the equation is still held, which implies the equation can be obtained by the mechanical energy conservation in the form $[d(\partial L/\partial \dot{q}_i)/dt - (\partial L/\partial q_i)]\dot{q}_i = 0$ shown in [22]. Whether this equation is equivalent or not to the general equations in mechanics, such as Newton's second law of motion, Lagrange's equation, or Hamilton's variational principle is left unresolved. For the last decade, there were several sporadic results reported on this issue although debates still exist [14, 24, 25].

Although in Carlson's work, the 'prescription equation', i.e., equation (7) in [14] to connect the rate of momentum with time $t$, with conservative forces were employed to formulate dynamical equations, it seems that his discussions indirectly used Newton's second law of motion in a conservative system. Until now there have not been systematic studies on how to start from the principle of energy conservation to formulate general dynamical equations in mechanics in a general sense.

On the other hand, it is fair to say that Hamilton's principle is the heart of physics, and definitely, it plays a significant role in leading scientists to go forward to the quantum world. However, it has been questioned in serval aspects [26, 27]. For example, based on variational concepts, the formulation of a dynamical equation or model is transformed to seek the correct action functional, in which Lagrange's equations govern the system's dynamics. However, the action is ambiguous and not directly measured, and its construction may only be justified postfactum by supplying experimentally verifiable equations of motion, and more importantly, there are phenomena beyond their scope for physical explanations, such as in a system with dissipations [26].

In our work, rather than starting from the dynamical equation to energy conservation principles, we start from the first principle of physics, i.e., the law of conservation of energy, to reconstruct the major results in Lagrangian and Hamiltonian mechanics, from an alternate point of view with energy and work quantities as required in Hamilton's principle. As will be seen, Lagrange's equation, Hamilton's canonical equations, and Hamilton-Jacobi's equations are reformulated from our proposed energy conservation methodology. The roadmap in our work for developing mechanics equations, such as Lagrange's equation, Hamilton's canonical equations, and Hamilton-Jacobi's equation using our methodology is shown in figure 1.

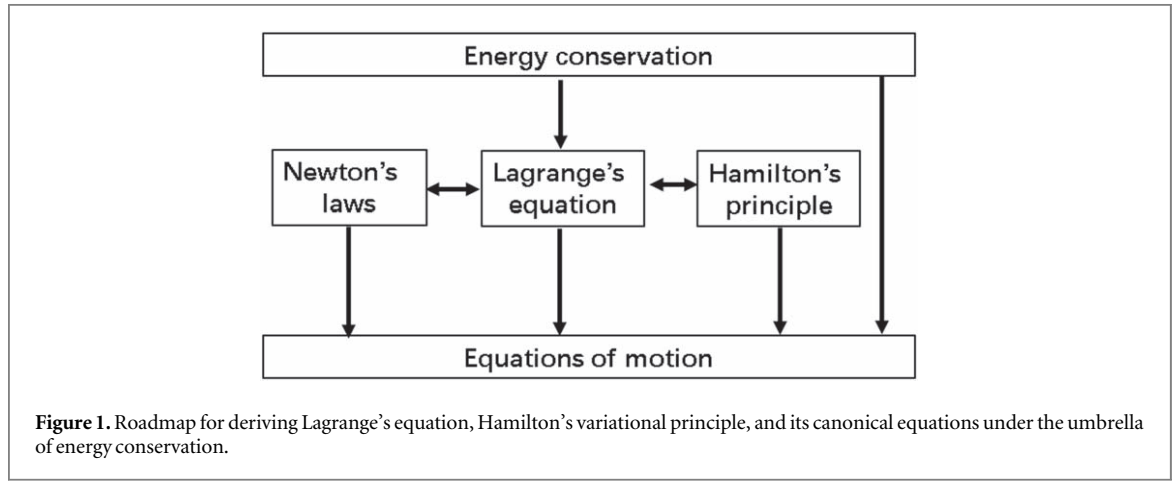

**Figure 1.** Roadmap for deriving Lagrange's equation, Hamilton's variational principle, and its canonical equations under the umbrella of energy conservation.

## 2. The basic equations in Hamiltonian mechanics

Some of the basic concepts and equations in Hamiltonian Mechanics are briefly recounted as they will be used in the coming-up discussions.

Following the system descriptions by Landau [1], consider $N$-multiple particles with $N$-masses and suppose there are $K$ constraints so that the number of degrees of freedom is $3N - K$, denoted by $M$ in a Cartesian coordinate. The spatial coordinates are defined as $r_\alpha$, with $\alpha = 1, 2, 3,...,3N$. Let the correspondent independent generalized coordinates be $q_i$, so that $i = 1, 2, 3,...,M$. The correspondent generalized velocity, generalized force, and generalized or conjugate momentum, are denoted as $\dot{q}_i$, $Q_i$, $p_i$, respectively, where $i = 1, 2, 3,...,M$. For this system, the kinetic and potential energy is written as $T = T(q_i, \dot{q}_i, t)$ and $U = U(q_i)$ in a generalized coordinate system, respectively. The basic concepts and principles in Hamiltonian mechanics are summarized as in the followings:

Lagrange's function is defined as

$$L = T(q_i, \dot{q}_i, t) - U(q_i), \tag{1a}$$

or simply, called Lagrangian, from which the generalized momentum is defined as [1, 28]

$$p_i = (\partial L/\partial \dot{q}_i). \tag{1b}$$

In a conservative system, from equation (1b) it is known that [1]

$$\dot{p}_i = d(\partial L/\partial \dot{q}_i)/dt. \tag{1c}$$

Lagrange's equation in a general case is written as in the following [16, 28]

$$d(\partial L/\partial \dot{q}_i)/dt - \partial L/\partial q_i - Q_i = 0, \tag{1d}$$

where $Q_i$ is a generalized force. The Hamiltonian is defined as [1, 16, 18, 28]

$$H = \dot{q}_i p_i - L, \tag{2a}$$

which is the Legendre's transformation through Lagrangian $L$. In the above formula and the following discussions, Einstein's summation convention is employed [29]. For example,

$$\dot{q}_i p_i = \sum_{i=1}^{M} \dot{q}_i p_i.$$

Furthermore, Hamilton's canonical equations can be derived from Lagrange's equations (1d) and (2a), and are written as [1, 18]

$$\dot{p}_i = -\partial H/\partial q_i, \tag{2b}$$

$$\dot{q}_i = \partial H/\partial p_i. \tag{2c}$$

However, more detailed derivation for general Hamilton's equations from the principle of least action with constraints of $\varphi(p_k, q_k) = 0$, with $k = 1, 2, 3,...,L < M$ are seen in [30, 31]. To focus on our main discussions, here we consider the situation without constraints between $p_k$ and $q_k$, i.e., the number of degrees of freedom is $M$, so that equations (2b) and (2c) are held.

### 3. From the principle of energy conservation to Lagrange's equation

The concepts of kinetic and potential energy and the principle of conservation of energy pre-date Newton's Laws, and in fact, they can be dated back to the fourteenth century [32]. Furthermore, for a while after Newton, the conservation of energy and Newton's Laws are thought to be independent and was later thought that the principle of energy conservation was a consequence of Newton's laws for certain special cases [32], i.e., the equations of motion yield a fundamental theorem called the law of the conservation of energy. On the contrary, in general, as is known, this principle at least is one of the backbones in physics. It can be applied to almost all science and practical areas. In this work, limited to mechanics, we will show how the dynamical equations of a system could be directly obtained from the law of the conservation of energy.

For every system, there exists an extensive, scalar state function $E$, called energy. If the system is isolated, the energy is constant, i.e., $E$ is a conserved observable. If the system is not isolated, then [7, 33]

$$dE/dt = P_w + P_q + P_c,$$

where $P_w$ is the power due to external forces acting on the system; $P_q$ is the power due to the heat transfer and $P_c$ is the power due to the matter transfer between the system and the exterior. The above equation is usually called the first law of thermodynamics, or the law of energy conservation.

In our discussions, heat and chemical effects are neglected, then the above equation is written as [34]

$$dE/dt = dW/dt, \tag{3a}$$

where $P_w = dW/dt$. $E$ is defined as the total energy of a system, and it includes kinetic and potential energy; while $W$ is the work done by external forces applied to the system, such as nonconservative forces, including dissipative forces, and so on. Generally, equation (3a) quantifies the relationship between mechanical energy and work done by external forces. It is this axiom that we use to formulate the existing mechanical equations.

To use the principle of energy conservation or equation (3a) to formulate Lagrange's equation and so on, one of the most important points in our work is how to incorporate Lagrangian $L$ or Hamiltonian $H$ of a system into equation (3a). As is known, Hamilton's function is seen as generalized energy, and usually, it is not equal to the total energy of a mechanical system, i.e., the sum of kinetic energy and potential energy. However, since Lagrangian is defined with kinetic minus potential energy, while Hamiltonian is defined with Legendre's transformation through Lagrangian, there must be a definite link among these functions associated with kinetic and potential energy.

The following is shown how to incorporate Lagrangian or Hamiltonian into equation (3a).

As is shown before, in a Cartesian coordinate system, there are $N$ particles with masses $m_\beta$, $\beta = 1, 2,...,N$, and their correspondent coordinate and velocity vectors are $\vec{r}_\beta$ and $\dot{\vec{r}}_\beta$ respectively, with $r_\alpha$, $\dot{r}_\alpha$, $\alpha = 1, 2,...,3N$ being their vector components respectively. The kinetic energy is written as (for example, in [1, 16, 35])

$$T(\dot{\vec{r}}_\beta) = T(\dot{r}_\alpha) = m_\beta \dot{\vec{r}}_\beta{}^2/2,$$

and the potential energy is denoted as $V = V(r_\alpha)$ (for example, in [1, 16, 35]). As is shown later, when generalized coordinates are introduced, Lagrangian $L$ might be explicitly time-dependent, i.e.,

$$L = L(r_\alpha, \dot{r}_\alpha) = L(q_i, \dot{q}_i, t).$$

According to equation (3a), the statement for the principle of energy conservation in mechanics is simplified into

$$d(T + V)/dt = dW/dt.$$

From our definitions, the full time-derivative of kinetic energy can be represented as

$$dT/dt = \dot{r}_\alpha d(\partial T/\partial \dot{r}_\alpha)/dt.$$

Using the above equations, the law of energy conservation in equation (3a) can be written as

$$[d(\partial T/\partial \dot{r}_\alpha)/dt + \partial V/\partial r_\alpha]\dot{r}_\alpha = dW/dt. \tag{3b}$$

Therefore, if we write the work done by forces $F_i$ per unit time $t$, that is, $dW/dt = F_\alpha \dot{r}_\alpha$ [34], equation (3b) can be rewritten as

$$[d(\partial T/\partial \dot{r}_\alpha)/dt + \partial V/\partial r_\alpha]\dot{r}_\alpha = F_\alpha \dot{r}_\alpha. \tag{3c}$$

Since potential energy is independent of $\dot{r}_\alpha$, while kinetic energy is independent of $r_\alpha$, it implies that $\partial V/\partial \dot{r}_\alpha$ and $\partial T/\partial r_\alpha$ vanish. Therefore, equation (3c) is shown as

$$[d(\partial L/\partial \dot{r}_\alpha)/dt - \partial L/\partial r_\alpha - F_\alpha]\dot{r}_\alpha = 0. \tag{3d}$$

Because the variable $\dot{r}_\alpha$ may not be independent if there exist constraints in $\dot{r}_\gamma$ or $r_\gamma$, $\gamma = 1, 2,...,K < 3N$, we could not drop off $\dot{r}_\alpha$ in the above equation to formulate Lagrange's equation. Nevertheless, generalized coordinate transformations are needed to transform $r_\alpha$ into $q_i$ so that $q_i$ are independent. To conduct generalized coordinate transformation, suppose

$$r_\alpha = r_\alpha(q_i, t), \ \alpha = 1, 2,...,3N; \ i = 1, 2,...,M,$$

where $q_i$ and $t$ are considered to be independent variables. In this case, Lagrangian is transformed from $L(r_\alpha, \dot{r}_\alpha)$ into $L(q_i, \dot{q}_i, t)$. We may conduct the following transformations that will be used later, i.e.,

$$\partial L/\partial q_i = (\partial L/\partial r_\alpha)(\partial r_\alpha/\partial q_i) + (\partial L/\partial \dot{r}_\alpha)(\partial \dot{r}_\alpha/\partial q_i). \tag{3e}$$

By using $\partial \dot{r}_\alpha/\partial \dot{q}_i = \partial r_\alpha/\partial q_i$, and $\partial r_\alpha/\partial \dot{q}_i = 0$ (for example, in [18, 22]), we have

$$\begin{aligned}\partial L/\partial \dot{q}_i &= (\partial L/\partial r_\alpha)(\partial r_\alpha/\partial \dot{q}_i) + (\partial L/\partial \dot{r}_\alpha)(\partial \dot{r}_\alpha/\partial \dot{q}_i) \\ &= (\partial L/\partial \dot{r}_\alpha)(\partial r_\alpha/\partial q_i),\end{aligned} \tag{3f}$$

and

$$\begin{aligned}d(\partial L/\partial \dot{q}_i)/dt = &[d(\partial L/\partial \dot{r}_\alpha)/dt](\partial r_\alpha/\partial q_i) \\ &+ (\partial L/\partial \dot{r}_\alpha)d(\partial r_\alpha/\partial q_i)/dt.\end{aligned} \tag{3g}$$

Since $r_\alpha = r_\alpha(q_i, t)$, the function $\partial r_\alpha/\partial q_i$ may contain variables $q_i$ and $t$, respectively. Therefore,

$$d(\partial r_\alpha/\partial q_i)/dt = [\partial(\partial r_\alpha/\partial q_i)/\partial q_j]\dot{q}_j + \partial^2 r_\alpha/\partial q_i \partial t.$$

Since $q_i$, $\dot{q}_i$ and $t$ are independent variables defined in generalized coordinate transformation, their partial differential order can be interchanged if $(\partial r_\alpha/\partial q_i)$ is continuous and differentiable. Therefore, the above equation can be simplified as [18]

$$d(\partial r_\alpha/\partial q_i)/dt = \partial[(\partial r_\alpha/\partial q_j)\dot{q}_j + \partial r_\alpha/\partial t]/\partial q_i.$$

By using

$$\dot{r}_\alpha = (\partial r_\alpha/\partial q_j)\dot{q}_j + \partial r_\alpha/\partial t, \tag{3h}$$

We have

$$d(\partial r_\alpha/\partial q_i)/dt = \partial \dot{r}_\alpha/\partial q_i$$

Using the above equation and equations (3e) and (3g), we have

$$\begin{aligned}&d(\partial L/\partial \dot{q}_i)/dt - \partial L/\partial q_i \\ =&[d(\partial L/\partial \dot{r}_\alpha)dt](\partial r_\alpha/\partial q_i) + (\partial L/\partial \dot{r}_\alpha)(\partial \dot{r}_\alpha/\partial q_i) \\ &-(\partial L/\partial r_\alpha)(\partial r_\alpha/\partial q_i) - (\partial L/\partial \dot{r}_\alpha)(\partial \dot{r}_\alpha/\partial q_i).\end{aligned}$$

Furthermore, the above equation can be simplified into

$$\begin{aligned}&d(\partial L/\partial \dot{q}_i)/dt - \partial L/\partial q_i \\ =&[d(\partial L/\partial \dot{r}_\alpha)dt - \partial L/\partial r_\alpha](\partial r_\alpha/\partial q_i).\end{aligned} \tag{3i}$$

Denote

$$H_\alpha = d(\partial L/\partial \dot{r}_\alpha)/dt - \partial L/\partial r_\alpha, \tag{3j}$$

and using equation (3h), equation (3d) can be rewritten as

$$\begin{aligned}&[H_\alpha(\partial r_\alpha/\partial q_i) - F_\alpha(\partial r_\alpha/\partial q_i)]\dot{q}_i \\ &+(H_\alpha - F_\alpha)(\partial r_\alpha/\partial t) = 0.\end{aligned} \tag{3k}$$

By using equation (3i), and the definition of a generalized force, $Q_i = F_\alpha(\partial r_\alpha/\partial q_i)$, the above equation is written as

$$\begin{aligned}&[d(\partial L/\partial \dot{q}_i)/dt - \partial L/\partial q_i - Q_i]\dot{q}_i \\ &+(H_\alpha - F_\alpha)(\partial r_\alpha/\partial t) = 0.\end{aligned} \tag{3l}$$

So far, we have incorporated Lagrangian into equation (3l), the law of the conservation of energy in a general mechanical system, and it will be used to formulate various dynamical equations later. Equation (3l) can be grouped in the following situations:

No. 1. As is known, if the system is conservative, then the total energy can be described with $H$, and the conservation of mechanical energy is written as

$$dH/dt = [d(\partial L/\partial \dot{q}_i)/dt - \partial L/\partial q_i]\dot{q}_i = 0. \tag{4a}$$

Equation (4a) is a special situation in equation (3l) with $Q_i = 0$ and $\partial r_\alpha/\partial t = 0$, respectively. The above equation, for example, was discussed in [16, 18, 22, 36].

No. 2. If there exist generalized forces in a mechanical system, the mechanical energy is no longer conserved itself, while the mechanical energy and work done by generalized forces on the system are still balanced in this case, which is given by using equation (3*l*) with $\partial r_\alpha / \partial t = 0$, i.e.,

$$dH/dt = Q_i \dot{q}_i, \quad i = 1, 2, ..., M. \tag{4b}$$

The physical meaning of the above equation is clear, i.e., the generalized energy rate with time is equal to the work done by generalized forces per unit time, which is a special case of equation (3*l*) subject to the condition of $\partial r_\alpha / \partial t = 0$.

No. 3. If there are external forces $F_\alpha$ applied to the system and $\partial r_\alpha / \partial t \neq 0$, the general case in equation (3*l*) should be taken into account. By using Legendre's transformation shown in equation (2*a*), we have

$$dH/dt = [d(\partial L/\partial \dot{q}_i)/dt - \partial L/\partial q_i]\dot{q}_i - \partial L/\partial t. \tag{4c}$$

Using the above equation, equation (3*l*) is written as

$$dH/dt + \partial L/\partial t - Q_i \dot{q}_i + (H_\alpha - F_\alpha)(\partial r_\alpha/\partial t) = 0. \tag{4d}$$

Equation (4*d*) is similar to the generalized energy theorem given by Cline [16]. If $\partial r_\alpha / \partial t = 0$, which is a scleronomic system, equation (4*d*) tells the exact generalized energy theorem [16] when all of the coordinates $q_i$ are independent. However, equation (4*d*) is given only based on the principle of energy conservation, while the existing generalized energy theorem was derived by using Lagrange's equation.

In the following, we will formulate the general dynamical equations by each of the situations stated in the previous sections.

### 3.1. Energy conservation in a conservative system with $H = H(q_i, \dot{q}_i)$

For a conservative dynamical system, the coordinate variables $r_\alpha$ may contain the time variable $t$ inexplicitly, that is, $\partial r_i / \partial t = 0$. Therefore, the law of conservation of mechanical energy is rewritten as in equation (4*a*), or

$$dE/dt = dH/dt = 0. \tag{5a}$$

The fact that the above equation holds is because, in a conservative system, the Hamiltonian is equal to the total mechanical energy. Therefore, $H$ in equation (2*a*), is the total energy of the system, and is characterized by using general coordinates and velocities, i.e., $H = H(q_i, \dot{q}_i)$ in configuration space, or $H = H(q_i, p_i)$ in phase space, where $q_i$, $\dot{q}_i$ and $q_i$, $p_i$ are two pairs of independent variables, respectively. Equation (5*a*) holds no matter whether it is in generalized or Cartesian coordinate systems. By using this mechanical energy conservation formula, Lagrange's equation will be derived. As mentioned before, $dH/dt$ can be expressed with Lagrangian, i.e.,

$$dH/dt = \dot{q}_i \dot{p}_i - (\partial L/\partial q_i)\dot{q}_i. \tag{5b}$$

By using equation (1*b*), the definition of generalized momentum, the above equation becomes

$$dH/dt = [d(\partial L/\partial \dot{q}_i)/dt - \partial L/\partial q_i]\dot{q}_i. \tag{5c}$$

As is assumed in a conservative system, the Lagrangian does not depend on time $t$ explicitly, that is to say, $L = L[q(t), \dot{q}(t)]$, and $\partial L/\partial t = 0$. Using equations (5*c*), (5*a*) can be written as

$$[d(\partial L/\partial \dot{q}_i)/dt - \partial L/\partial q_i]\dot{q}_i = 0. \tag{5d}$$

The above equation was also given in [22], where Pars claimed that Lagrange's equation could not be derived from equation (5*d*) because the choice $\dot{q}_i$ is not arbitrary. However, we argue that since $q_i$ and $\dot{q}_i$ are independent variables of Lagrangian mathematically [1], there are no constraints on them. This could be seen clearly by multiplying both sides of equation (5*c*) with time deferential $dt$, and mathematically we obtain the following equation, i.e.,

$$[d(\partial L/\partial \dot{q}_i)/dt - \partial L/\partial q_i]dq_i = 0. \tag{5e}$$

As we know, since $q_i$, $\dot{q}_i$, $t$ are independent variables, $dq_i$ and $dt$ are also independent variables, respectively. Now, we prove that the above equation implies Lagrange's equation using the method of proof by contradiction.

Denote $C_i = [d(\partial L/\partial \dot{q}_i)/dt - \partial L/\partial q_i]$, where $C_i$ may be the functions of $q_i$, $\dot{q}_i$, $t$, and are the correspondent coefficients of independent variables $dq_i$. Equation (5*e*) is shortly written as

$$C_i dq_i = 0, \quad i = 1, 2, ..., M. \tag{5f}$$

Suppose $C_l$ are some of the coefficients or all of $C_i$ for the correspondent independent variable $dq_l$ and they are not equal to zero, that is, $C_l(t) \neq 0$, $l \in \{1, 2, 3, ..., K \leqslant M\}$, and dropping off the zero terms in equation (5*f*), yields

$$C_l dq_l = 0.$$

From the above equation, we know that the correspondent $dq_l$ are dependent since $C_l \neq 0$. However, this is against our assumption that, $dq_i$ are totally independent. Therefore, from equations (5*e*) or (5*f*), we know that all of the coefficients $C_i$ must vanish, which leads to

$$d(\partial L/\partial \dot{q}_i)/dt - \partial L/\partial q_i = 0. \tag{6}$$

Another way to derive Lagrange's equation from equation (5d) is that since $dq_i$ are independent of each other, the quantities $dq_i$ are accordingly freely selectable, we could, for example put to zero all of $dq_i$ except for one, and that has the consequence that in equation (5d) not only the sum but even each summand vanishes, which is equivalent to one dimension problem, so that we obtain equation (6) [18]. There are also other ways in dropping off $dq_i$ that was used in [18, 35].

Equation (6) is the exact Lagrange's equation that we want from the law of conservation of energy and is in an agreement with equation (1d) with $Q_i = 0$ for a conservative system.

### 3.2. Energy conservation in a nonconservative system $H = H(q_i, \dot{q}_i)$

In the second case, $\partial r_\alpha/\partial t = 0$, and generalized forces $Q_i \neq 0$. We start from equation (4b). With the help of $H$, Lagrangian $L$ will be identified by using Legendre's transformation, and the law of the conservation of energy in equation (4b), can be quantified with the following formula

$$[d(\partial L/\partial \dot{q}_i)/dt - (\partial L/\partial q_i)]\dot{q}_i = Q_i\dot{q}_i, \tag{7a}$$

or

$$[d(\partial L/\partial \dot{q}_i)/dt - (\partial L/\partial q_i) - Q_i]dq_i = 0. \tag{7b}$$

Since $dq_i$ are independent variables, following the previous procedures, we obtain

$$d(\partial L/\partial \dot{q}_i)/dt - (\partial L/\partial q_i) - Q_i = 0. \tag{8}$$

The above equation is just the Lagrange's equation formulated from the law of the conservation of energy, seen in equation (1d).

### 3.3. Energy conservation in a nonconservative system with $H = H(q_i, \dot{q}_i, t)$

In the third case, Lagrangian and Hamiltonian contain an explicit variable of time $t$, which means, in addition to $Q_i \neq 0$, $\partial r_\alpha/\partial t \neq 0$. From equation (3l), we have

$$[d(\partial L/\partial \dot{q}_i)/dt - \partial L/\partial q_i - Q_i]dq_i$$
$$+[(H_\alpha - E_\alpha)(\partial r_\alpha/\partial t)]dt = 0. \tag{9a}$$

Following the previous procedures, by considering the independent and arbitrary choices of $dq_i$ and $dt$, we have

$$d(\partial L/\partial \dot{q}_i)/dt - \partial L/\partial q_i - Q_i = 0, \tag{9b}$$

and

$$(H_\alpha - E_\alpha)(\partial r_\alpha/\partial t) = 0. \tag{9c}$$

Equation (9b) is the general Lagrange's equation, seen in equation (1d). Since $\partial r_\alpha/\partial t$ might be dependent, the value of $H_\alpha - E_\alpha$ might not be equal to zero in equation (9c). It seems that equation (9c) tells that there are some additional constraints imposed by the law of the conservation of energy, and in Lagrangian and Hamiltonian mechanics, there are no constraint descriptions, such as shown in equation (9c). However, from our later discussions, it is show that equation (9c) will not give any additional constraints than equation (9b), and it will be seen that equation (9b) is equivalent to equation (9c).

## 4. From the principle of energy conservation to Hamilton's canonical equations

Suppose that Lagrangian is $L = L(q_i, \dot{q}_i, t)$, in which the time may be an explicit variable, while $H = H(q_i, p_i, t)$ in phase space $(q_i, p_i)$, also a function of explicit variable $t$ in every system, it is known that [18, 35, 36]

$$dH/dt = (\partial H/\partial q_i)\dot{q}_i + (\partial H/\partial p_i)\dot{p}_i + \partial H/\partial t. \tag{10a}$$

Also, from equation (4c), we have

$$dH/dt = \dot{p}_i\dot{q}_i - (\partial L/\partial q_i)\dot{q}_i - \partial L/\partial t. \tag{10b}$$

Comparing equations (10a) and (10b) with each other, yields

$$(\partial H/\partial q_i + \partial L/\partial q_i)\dot{q}_i + (\dot{q}_i - \partial H/\partial p_i)\dot{p}_i$$
$$+\partial H/\partial t + \partial L/\partial t = 0. \tag{11a}$$

Multiplying the above equation by $dt$, yields

$$(\partial H/\partial q_i + \partial L/\partial q_i)dq_i + (\dot{q}_i - H/\partial p_i)dp_i$$
$$+(\partial H/\partial t + \partial L/\partial t)dt = 0. \tag{11b}$$

Following the previous procedures, considering $dq_i$, $dp_i$ and $dt$ are independent and their choices are arbitrary in phase space, we have

$$\begin{cases} \dot{q}_i = \partial H/\partial p_i, \\ \partial H/\partial q_i = -\partial L/\partial q_i, \\ \partial H/\partial t = -\partial L/\partial t. \end{cases} \tag{12}$$

Our treatments on equations (10a) and (10b) for equation (12) are the same as those used in [18, 35]. On the other hand, by using equations (1c) and (3l), we know

$$(\dot{p}_i - \partial L/\partial q_i - Q_i)\dot{q}_i + (H_\alpha - F_\alpha)(\partial r_\alpha/\partial t) = 0, \tag{13a}$$

Multiplying both sides of equation (13a) by $dt$, yields

$$(\dot{p}_i - \partial L/\partial q_i - Q_i)dq_i + (H_\alpha - F_\alpha)(\partial r_\alpha/\partial t)dt = 0. \tag{13b}$$

Since $dq_i$ and $dt$ are independent and their choices are arbitrary, following the previous procedures, we obtain that

$$\begin{cases} \dot{p}_i = \partial L/\partial q_i + Q_i, \\ (H_\alpha - F_\alpha)(\partial r_\alpha/\partial t) = 0. \end{cases} \tag{14}$$

From equations (12) and (14), a group of equations is obtained as in the followings

$$\begin{cases} \dot{q}_i = \partial H/\partial p_i, \\ \dot{p}_i = -\partial H/\partial q_i + Q_i, \\ \partial L/\partial t = -\partial H/\partial t, \\ (H_\alpha - F_\alpha)(\partial r_\alpha/\partial t) = 0. \end{cases} \tag{15a}$$

When $Q_i = 0$, the above equations become

$$\begin{cases} \dot{q}_i = \partial H/\partial p_i, \\ \dot{p}_i = -\partial H/\partial q_i, \\ \partial L/\partial t = -\partial H/\partial t, \\ (H_\alpha - F_\alpha)(\partial r_\alpha/\partial t) = 0. \end{cases} \tag{15b}$$

The above equations contain Hamilton's canonical equations. Not only can we derive Hamilton's canonical equations, but also, it seems that we obtain an additional equation $(H_\alpha - F_\alpha)(\partial r_\alpha/\partial t) = 0$ to quantify $r_\alpha$ and $F_\alpha$ so that the motion track $r_\alpha$ are subject to energy conservation shown in equation (3a), or more specifically in equation (3l), which is equation (9c). Logically, equation (9c) or the last equation in equation group (15b) must be agreed with Lagrange's equation, or at least it should not be contradicted with it.

Now, let us examine equation (9c) or the last one in equation group (15b). Substituting equation (3j) into equation (9c), yields

$$[d(\partial L/\partial \dot{r}_\alpha)/dt - \partial L/\partial r_\alpha - F_\alpha](\partial r_\alpha/\partial t) = 0. \tag{16a}$$

Recalling equations (3i) and (9b) and the definition of generalized force, we have

$$[d(\partial L/\partial \dot{r}_\alpha)/dt - \partial L/\partial r_\alpha - F_\alpha](\partial r_\alpha/\partial q_i) = 0, \tag{16b}$$

which is Lagrange's equation derived from the law of the conservation of energy. Multiplying equations (16a) and (16b) with $dt$ and $dq_i$, respectively, and then adding these equations together yields

$$[d(\partial L/\partial \dot{r}_\alpha)/dt - \partial L/\partial r_\alpha - F_\alpha]dr_\alpha = 0, \tag{16c}$$

Where $dr_\alpha = (\partial r_\alpha/\partial q_i)dq_i + (\partial r_\alpha/\partial t)dt$. The above equation is equivalent to equation (3d). If a scleronomic condition is considered, i.e., the time variable is 'frozen' temporally, then $dr_\alpha = (\partial r_\alpha/\partial q_i)dq_i + (\partial r_\alpha/\partial t)dt = (\partial r_\alpha/\partial q_i)dq_i$. In this case $(\partial r_\alpha/\partial q_i)dq_i = (\partial r_\alpha/\partial q_i)\delta q_i$ as $q_i$ and $t$ are independent mathematically. Hence,

$$dr_\alpha = \delta r_\alpha = (\partial r_\alpha/\partial q_i)\delta q_i,$$

and the above equation defines the 'virtual displacement' by 'freezing' time. In this case, equation (16c) becomes

$$[d(\partial L/\partial \dot{r}_\alpha)/dt - \partial L/\partial r_\alpha - F_\alpha]\delta r_\alpha = 0. \tag{16d}$$

The above equation is exactly the general dynamical equation, or D'Alembert's principle[18, 36].

On the other hand, since $q_i$ and $t$ are independent mathematically, we may set $q_i$ be 'frozen' temporally, and in this case, we denote the following variables as

$$s_\alpha = r_\alpha = r_\alpha(q_i, t)|_{q_i},$$ (17a)

$$\dot{s}_\alpha = \dot{r}_\alpha = d(r_\alpha|_{q_i})/dt = \partial r_\alpha/\partial t,$$ (17b)

$$\delta s_\alpha = dr_\alpha = (dr_\alpha|_{q_i}) = (\partial r_\alpha/\partial t)\delta t,$$ (17c)

where $r_\alpha|_{q_i}$ means the variable $q_i$ in $r_\alpha$ can be taken as a temporal constant. Equation (17c) defines the 'virtual displacement' by 'freezing' $q_i$. Therefore, the above representation means the variable transformation is conducted from $(r_\alpha, \dot{r}_\alpha)$ to $(s_\alpha, \dot{s}_\alpha, t)$. Equation (16a) is written as

$$[d(\partial L/\partial \dot{s}_\alpha)/dt - \partial L/\partial s_\alpha - F_\alpha]\delta s_\alpha = 0.$$ (17d)

where $L(r_\alpha, \dot{r}_\alpha) = L(s_\alpha, \dot{s}_\alpha, t)$. In the above equation, since $s_\alpha = s_\alpha(t)$ as $q_i = q_i(t)$. $s_\alpha = s_\alpha(t)$ doesn't make any difference from $q_i = q_i(t)$ except that $q_i$ is independent. Finally, it turns out that equation (17d) is identical with equation (16d), i.e., D'Alembert's principle.

From the law of the conservation of energy, mathematically there are two kinds of D'Alembert's principle. The first one is the original, with time-variable being temporally 'frozen' and $dr_\alpha = (\partial r_\alpha/\partial q_i)\delta q_i = \delta r_\alpha$; while the second one, with the generalized coordinate $q_i$ being temporally 'frozen' and $dr_\alpha = (\partial r_\alpha/\partial t)\delta t = \delta r_\alpha$. Both of those assumptions give the same results. All of those are derived from the law of the conservation of energy. Hence, equation (9c) or equation (17d) doesn't provide any more information than equation (16d) because they are equivalent to each other. Hence, equation (9c) will be dropped off during our later discussions.

The next step is to prove Hamilton's principle by using the derivatives from the principle of energy conservation through equation (3l).

Since equation (15a) or Hamilton's equations is the consequences of the law of energy conservation, Hamilton's principle in a nonconservative system [5, 9] could be proved. The principle states that, if $I$ is defined as

$$I = \int_{t_1}^{t_2} L dt + \int_{t_1}^{t_2} W_e dt,$$ (18a)

then

$$\delta I = \int_{t_1}^{t_2} \delta L dt + \int_{t_1}^{t_2} \delta W_e dt = 0,$$ (18b)

gives the dynamical equation of a nonconservative system, where $\delta W_e$ is the work done by generalized forces $Q_i$ along the virtual displacement $\delta q_i$.

Using the Legendre's transform, the Lagrangian $L$ is represented by Hamiltonian $H$. Hence,

$$\delta L + \delta W_e = (\dot{q}_i - \partial H/\partial p_i)\delta p_i + p_i\delta\dot{q}_i + (Q_i - \partial H/\partial q_i)\delta q_i.$$ (18c)

Furthermore,

$$\delta I = \int_{t_1}^{t_2} p_i\delta\dot{q}_i dt + \int_{t_1}^{t_2} [(\dot{q}_i - \partial H/\partial p_i)\delta p_i + (Q_i - \partial H/\partial q_i)\delta q_i] dt.$$ (18d)

Since

$$\int_{t_1}^{t_2} p_i\delta\dot{q}_i dt = p_i\delta q_i|_{\delta q_i(t_1)}^{\delta q_i(t_2)} - \int_{t_1}^{t_2} \dot{p}_i\delta q_i dt,$$ (18e)

by using the same time-terminal conditions as given in Hamilton's variational principle, i.e., $\delta q_i(t_1) = \delta q_i(t_2) = 0$ in equations (18e), (18d) can be simplified as

$$\delta I = \int_{t_1}^{t_2} [(Q_i - \partial H/\partial q_i - \dot{p}_i)\delta q_i + (\dot{q}_i - \partial H/\partial p_i)\delta p_i] dt.$$ (19)

Substituting the first and second equations in equation group (15a) into equation (19), yields equation (18b).

So far, it is found that the principle of least action or Hamilton's variational principle is a consequence of the law of the conservation of energy shown in equations (3a) or equation (3l). This is also the physics essence of Hamilton's variational principle.

Another version of the principle of least action was derived that the distinguishing feature of the theory is that the varied motion is subject throughout to the restriction that the total energy remains constant. Naturally, this theory is also a derivative of the principle of energy conservation [18, 22].

## 5. From energy conservation to Hamilton-Jacobi's equation

By using equation (3*l*) with its generalized forces $Q_i$ being zero, and dropping off the second term on the left-hand side of the equation as discussed previous sections, we have

$$[d(\partial L/\partial \dot{q}_i)/dt - \partial L/\partial q_i]\dot{q}_i = 0, \tag{20}$$

We will show that we still can formulate Hamilton-Jacob's equation from equation (20), which means that equation (20) is equivalent to Lagrange's equation and Hamilton's equations.

As is known, the action is defined in the form of [1, 18, 36]

$$S = S(q_i, t) = \int_0^t L(q_i, \dot{q}_i, t)dt. \tag{21a}$$

Therefore, $dS/dt = L$, and

$$\partial(dS/dt)/\partial t = \partial L/\partial t. \tag{21b}$$

Taking the full derivative of $S$ with respect to $t$, we have

$$dS/dt = (\partial S/\partial q_i)\dot{q}_i + \partial S/\partial t. \tag{22}$$

From the following derivation, we will see that, even as Pars claimed, it would be impossible to derive the Lagrange's equation from equation (20). However, we will still prove that equation (20) is equivalent to Hamilton-Jacobi's equation, which is another expression of general dynamical equations.

Since in this case $\partial L/\partial t \neq 0$, by using equation (20), we have the following formula

$$dH/dt + \partial L/\partial t = 0. \tag{23}$$

Using equations (21*b*) and (23), we have

$$dH/dt + \partial(dS/dt)/\partial t = 0. \tag{24a}$$

From equation (22), the second term on the left-hand side of equation (24*a*) can be written as

$$\partial(dS/dt)/\partial t = \partial[(\partial S/\partial q_i)\dot{q}_i]/\partial t + \partial^2 S/\partial t^2. \tag{24b}$$

Since $q_i, \dot{q}_i, t$ are defined as independent variables, and $S$ is the continuous and differentiable for $q_i$ and $t$, interchanging the partial derivative sequences will not change the partial derivative values, which means

$$\partial(dS/dt)/\partial t = [\partial^2 S/(\partial q_i \partial t)]\dot{q}_i + \partial^2 S/\partial t^2. \tag{25a}$$

Also, from the definition of the action $S$ in equation (21*a*), we have

$$d(\partial S/\partial t)/dt = [\partial^2 S/(\partial t \partial q_i)]\dot{q}_i + \partial^2 S/\partial t^2. \tag{25b}$$

Comparing equations (25*a*) and (25*b*) yields

$$d(\partial S/\partial t)/dt = \partial(dS/dt)/\partial t. \tag{26}$$

Therefore, from equations (24*a*) and (26), we have

$$dH/dt + d(\partial S/\partial t)/dt = 0, \tag{27a}$$

or

$$d(H + \partial S/\partial t)/dt = 0, \tag{27b}$$

which leads to

$$H + \partial S/\partial t = Const. \tag{28}$$

Taking the above constant Const to be zero, we obtain the Hamilton-Jacobi's equation,

$$H + \partial S/\partial t = 0. \tag{29}$$

It is safe to claim that equation (20) is equivalent to equation (29), which means equation (20), is equivalent to Lagrange's equation $d(\partial L/\partial \dot{q}_i)/dt - \partial L/\partial q_i = 0$.

So far, Lagrange's equation, Hamilton's canonical equations and Hamilton's variational principle, Hamilton-Jacobi's equation, as well as D'Alembert's principle are formulated only based on the principle of energy conservation, or the Law of conservation of energy. What is more is that all of the other conservation laws could be derived at some special conditions, including the laws of conservation of momentum and angular momentum. In a word, the law of conservation of energy is at least the root of mechanics.

## 6. Applications in fluid mechanics

In this section, an example is given to show how to use the law of the conservation of energy to formulate fluid dynamical equations that were usually done by using Newton's second law of motion or momentum theorem. The equation of energy conservation in fluid mechanics is written as [37, 38]

$$
d\left[\iiint_{\Omega(t)} (\rho \dot{u}_i \dot{u}_i/2 + \rho e)\, dv\right]/dt
$$
$$
= \oiint_{\partial\Omega(t)} T_i \dot{u}_i ds + \iiint_{\Omega(t)} \rho g_i \dot{u}_i dv - \oiint_{\partial\Omega(t)} q_i n_i ds. \tag{30}
$$

In the above equation, $\rho$ is material density; $u_i$ is the particle displacement component of the material, and the velocity is $\dot{u}_i = du_i/dt$; $e$ is the mass potential energy density; $g_i$ is the mass body force; $q_i$ is energy flux component; $n_i$ is the directional cosines of a unit vector normal to the surface element $ds$; the stress components on the surface $ds$ can be written as $T_i = \sigma_{ji} l_j$, $i, j = 1, 2, 3$ [5].

   The term on the left-hand side of equation (30) is the total energy included in a any given volume $\Omega$ with an enclosed surface $\partial\Omega$, and it consists of kinetic energy and potential energy, i.e., $\rho \dot{u}_i \dot{u}_i/2$ is kinetic energy density, while $\rho e$ is the potential energy density. The first term on the right-hand side of equation (30) is the work rate with time done by the stresses applied on the surface; the second term, the work rate with time done by the body force in the volume of interest; and the last one, the power flux transformed out of the volume passing through the surface. Since the energy flux goes out from the surface, its contribution to the system would be negative. The volume and the surface are changed with respect to time, so we consider them to be the functions of time.

   Since the integral volume $\Omega$ is an arbitrary choice, equation (30) can be easily written in a differential form by using Gauss's theorem, and converting the enclosed surface integral into the volume integral [37], i.e.,

$$
d(\rho \dot{u}_i \dot{u}_i/2 + \rho e)/dt = (\sigma_{ji} \dot{u}_i)_{,j} + \rho g_i \dot{u}_i - q_{i,i}. \tag{31}
$$

Substituting $(\sigma_{ji} \dot{u}_i)_{,j} = \sigma_{ji,j} \dot{u}_i + \sigma_{ji} \dot{u}_{i,j}$ into the above equation and rearranging it, yields

$$
(\rho \ddot{u}_i - \sigma_{ji,j} - \rho g_i) \dot{u}_i + \rho \dot{e} - \sigma_{ji} \dot{u}_{i,j} + q_{i,i} = 0,
$$

or multiplying both sides of the above equations with time differential element $dt$, yields

$$
(\rho \ddot{u}_i - \sigma_{ji,j} - \rho g_i) du_i + (\rho \dot{e} - \sigma_{ji} \dot{u}_{i,j} + q_{i,i}) dt = 0. \tag{32}
$$

Since $du_i$ and $dt$ in the above equation are arbitrary and independent, their corresponding coefficients must vanish, i.e.,

$$
\rho \ddot{u}_i - \sigma_{ji,j} - \rho g_i = 0, \tag{33a}
$$

and

$$
\rho \dot{e} - \sigma_{ji} \dot{u}_{i,j} + q_{i,i} = 0. \tag{33b}
$$

Equation (33a) is known as the Cauchy's first law of motion with heat effect, which is usually derived by using Newton's second law of motion or momentum balance [37]. Equation (33b) implies the constitutive relations for the material of interest, and it is equation (2.119) in [37], and more physical meanings are stated by Spurk and Aksel.

   We now just take a simple example to show the physical meaning of equation (33b). In perfectly elastic material, the heat effect may be neglected in elastodynamics. Therefore, equation (33b) can be written as

$$
\rho \dot{e} - \sigma_{ji} \dot{u}_{i,j} = 0.
$$

According to the definition, $\rho e = U_\rho$ can be seen as elastic potential energy density in the continuous material of interest. Furthermore, the above equation is written as

$$
\dot{U}_\rho - \sigma_{ji} \dot{u}_{i,j} = 0.
$$

On the other hand, the generalized Hook's law for anisotropic continous system can be written as $\sigma_{ji} = C_{jikl} e_{kl}$ [9], where $e_{kl} = (u_{k,l} + u_{l,k})/2$. Apparently, according to the relationship between linear strains and displacements, Hooke's law can be rewritten as $\sigma_{ji} = C_{jikl} u_{k,l}$ by using $C_{jikl} = C_{jilk}$, so that the strain energy density rate with time is rewritten as

$$
\dot{U}_\rho - C_{jikl} u_{k,l} \dot{u}_{i,j} = 0,
$$

or

$$
d(U_\rho - C_{jikl} u_{k,l} u_{i,j}/2)/dt = 0,
$$

which identifies the elastic potential energy density, i.e.,

$$U_\rho = C_{jikl} u_{k,l} u_{i,j} / 2. \tag{34}$$

Now, look back at equations (33*a*) and (33*b*), we know that the first equation gives the dynamical equation of motion of material particles, which is the Cauchy's first law of motion; while the second, the constitutive relations of the medium. All of these equations are formulated only by the law of energy conservation.

## 7. Conclusion and discussions

We proposed a novel methodology in reformulating Lagrange's equations, Hamilton's variational principle, Hamilton's canonical equations, and Hamilton-Jacobi's equation under the umbrella of energy conservation. It is pointed out that, the reasons why Lagrange's equation and Hamilton's principle are held in a dynamical system is because of the energy conservation. Hamilton's variational principle is a mathematical description of a system, and the physics essence behind the existing mechanics' principles is energy conservation. Therefore, the principle of least action or Hamilton's variational principle is also the consequence of the energy conservation, i.e., the principle of energy conservation governs the dynamical system's behaviors. The advantages of our mechanical formulations are that the framework is clear and easy to understand, and even no variational concepts and principles are introduced. Our methodology could be easily extended to electrodynamics, fluid dynamics, and elastodynamics, and so on.

## Acknowledgments

The authors would like to thank our colleagues at Institute of Acoustics in Chinese Academy of Sciences (IACAS), Dr. Xiumei Zhang, Dr. Xiao He, Dr. Hao Chen, Dr. Dehua Chen, Dr. Chang Su, Dr. Pengfei Wu, Dr. Weijun Lin, Dr. Lixin Bai, Dr. Hanyin Cui, Dr. Delong Xu for their critical comments and some of profound discussions, and Dr. Tingting Liu for her checks in some of mathematical derivations. The authors also would like to thank Prof. Hailan Zhang, Prof. Jing Tian, and Prof. Xiaomin Wang at IACAS for their helpful discussions and suggestions. The authors are so grateful for the critical comments from two anonymous reviewers that stimulated the improvement of the work. Last but not least, the authors would like to thank Prof. Ning Wang from Ocean University of China, Prof. Mingxiang Pan from Institute of Physics in Chinese Academy of Sciences, and Prof. Hengshan Hu from Harbin Institute of Technology in China for their inspirations and suggestions during the preparations for this paper.

The work has been partially supported by the National Natural Science Foundation of China with Grant No. 11974018.

## Data availability statement

All data that support the findings of this study are included within the article (and any supplementary files).

## ORCID iDs

Yinqiu Zhou  https://orcid.org/0000-0002-5646-7094
Xiuming Wang  https://orcid.org/0000-0002-0499-7827